\documentclass[natbib]{nature_fig}

\usepackage{sansmath}
\usepackage{amssymb}
\usepackage{color}

\usepackage{graphicx}
\usepackage{epsfig}
\usepackage{upgreek}
\usepackage{xcolor}
\usepackage{lineno}
\usepackage{hyperref}
\usepackage{rotating}
\usepackage{textcomp}
\usepackage{threeparttable}
\usepackage{booktabs}
\usepackage{caption}
\usepackage[caption = false]{subfig}
\usepackage{ulem}
\usepackage{amsmath}

\newcommand{\lsi}{LS~I~+61$^{\circ}$~303}
\newcommand{\fast}{{\it FAST}}

\title{Radio pulsations from a neutron star within the gamma-ray binary \lsi{}}
\author{Shan-Shan~Weng$^{1*\dag}$, Lei~Qian$^{2,3*}$, Bo-Jun~Wang$^{4,5,2*}$,
D. F. Torres$^{6,7,8\dag}$, A. Papitto$^{9}$, Peng~Jiang$^{2,3}$, Renxin~Xu$^{4,5}$, Jian~Li$^{10,11}$,
Jing-Zhi~Yan$^{12}$, Qing-Zhong~Liu$^{12}$, Ming-Yu~Ge$^{13}$, Qi-Rong~Yuan$^{1}$}

\newcommand{\araa}{Annu. Rev. Astron. Astrophys.}   
\newcommand{\aj}{Astron. J.}   
\newcommand{\apj}{Astrophys. J.}   
\newcommand{\apjl}{Astrophys. J. Lett.}   
\newcommand{\aap}{Astron. Astrophys.}   
\newcommand{\mnras}{Mon. Not. R. Astron. Soc.}   
\newcommand{\nat}{Nature} 

\begin{document}

\maketitle

\begin{affiliations}
 \item Department of Physics and Institute of Theoretical Physics, Nanjing Normal University, Nanjing 210023, China
 \item National Astronomical Observatories, Chinese Academy of Sciences, 20A Datun Road, Chaoyang District, Beijing, 100101, China
 \item CAS Key Laboratory of FAST, National Astronomical Observatories, Chinese Academy of Sciences, Beijing 100101, China
 \item Kavli Institute for Astronomy and Astrophysics, Peking University, Beijing 100871, China
 \item Department of Astronomy, School of Physics, Peking University, Beijing 100871, China
\item  Instituci\'o Catalana de Recerca i Estudis Avan\c cats (ICREA), 08010 Barcelona, Spain
 \item Institute of Space Sciences (ICE, CSIC), Campus UAB, Carrer de Can Magrans s/n, 08193 Barcelona, Spain 
\item Institut d'Estudis Espacials de Catalunya (IEEC), 08034 Barcelona, Spain 
\item  INAF-Osservatorio Astronomico di Roma (OAR) via di Frascati, 33 I-00044, Monte Porzio Catone, Italy
\item  CAS Key Laboratory for Research in Galaxies and Cosmology, Department of Astronomy, University of Science and Technology of China, Hefei 230026, China
\item School of Astronomy and Space Science, University of Science and Technology of China, Hefei 230026, China

\item  Key Laboratory of Dark Matter and Space Astronomy, Purple Mountain Observatory, Chinese Academy of Sciences, Nanjing 210023, China
\item Key Laboratory of Particle Astrophysics, Institute of High Energy Physics, Chinese Academy of Sciences, Beijing 100049, China

  \item[$^{*}$] Co-First Authors
  \item[$^{\dag}$] Corresponding Authors: wengss@njnu.edu.cn, dtorres@ice.csic.es

\end{affiliations}

\begin{abstract}

\lsi\ is one of the rare gamma-ray binaries\cite{dubus2013}, emitting most of their luminosity in photons with energies beyond 100 MeV\cite{abdo2009}.  The $\sim$26.5 d orbital period is clearly detected  at many wavelengths\cite{gregory2002,albert2008,abdo2009}. Additional aspects of its multi-frequency behavior  make it the most interesting example of the class. The morphology of high-resolution radio images changes with orbital phase displaying a cometary tail pointing away from the high-mass star\cite{Dhawan2006}. \lsi\ also shows superorbital variability \cite{gregory2002,Chernyakova2012,li2012,Ackermann2013,Ahnen2016}. A couple of energetic ($\sim 10^{37}$ erg s$^{-1}$), short, magnetar-like bursts have been plausibly ascribed to it\cite{depasquale2008,barthelmy2008,Arjonilla2009,Torres2012}. \lsi's phenomenology has been put under theoretical scrutiny for decades, but the lack of certainty regarding the nature of the compact object in the binary has prevented  advancing our understanding of the source. Here, using observations done with the Five-hundred-meter Aperture Spherical radio Telescope (\fast{}), we report on the existence of transient radio pulsations from the direction of \lsi. We find a period $P=269.15508 \pm 0.00016$ ms at a significance of $> 20\sigma$. This is the first evidence for pulsations from this source at any frequency, and strongly argues for the existence of a rotating neutron star in \lsi.

\end{abstract}


\lsi\ locates at a distance of $2.65\pm0.09$ kpc\cite{Lindegren2021} and contains a compact object orbiting a rapidly-rotating B0Ve star every 26.5 days\cite{gregory2002,casares2005}. The dynamical mass of the compact object is between 1 and 4 $M_{\odot}$; thus, just from dynamical arguments, it could either be a neutron star or a low-mass black hole\cite{casares2005,grundstrom2007,aragona2009}. Models involving an accreting black hole launching a relativistic jet (a microquasar) see  e.g.,\cite{massi2004}; a rotationally-powered neutron star emitting a relativistic wind of particles in interaction with the stellar wind of the companion, see e.g.,\cite{maraschi1981}, and a neutron star alternating between a rotationally-powered emission and propeller ejection of the mass lost by the companion, see e.g.,\cite{zamanov1995}, have been proposed to explain the multi-frequency phenomenology. Despite modern incarnations of these models were able to provide a framework where to interpret the growing number of observations,  the fact that they are based on dissimilar compact object scenarios stagnated progress.
%


Prior deep searches in the radio\cite{mcswain2011,canellas2012}, X-ray\cite{rea2010}, and gamma-ray band\cite{hadasch2012} were not successful in finding pulsations. This is in fact not surprising: free-free absorption -which may have a complex temporal behavior according to binary  conditions- can easily wash out the pulses in the radio band, see e.g.,\cite{Zdziarski2010}.  Also, the radio cone of emission may altogether point in a different direction from Earth. In X-rays, the imposed pulsed fraction upper limit of $\sim$10\%  (at 3$\sigma$ confidence level)  could well be larger than the actual pulsed fraction of the source, as is the case for other pulsars. In fact, only a few dozen pulsars out of the $\sim 300$ detected in gamma rays and the $\sim 3000$ in radio have non-thermal X-ray pulsations detected. Finally, in gamma-rays, \lsi\ lies in a complex and populated region, and not only the diffuse background, but the likely origin of at least part of the GeV emission beyond the magnetosphere of the putative pulsar may preclude detecting pulses. Additionally,  the relatively large uncertainty in the orbital parameters  reduces the sensitivity of blind searches across all frequencies when long integration times are needed\cite{caliandro2012}. The best chance to ever detect pulsations from \lsi\ was to try observing at a large radio sensitivity in the orbital region where the free-free absorption effect due to the stellar wind (or disk) would naturally be the lowest \cite{canellas2012}. 


\fast{} is the largest single-aperture radio telescope located in a naturally deep and round karst depression in Southwest China's Guizhou province (see {\bf Methods}). \fast{} executed four observations towards \lsi, with a total exposure time of $\sim 10.2$ hours (see Table \ref{tab:obs}): one at the orbital phase of $\sim 0.07$ and three around the orbital phase of $\sim 0.6$. The zero of orbital phase of \lsi\ is defined at MJD$_{0} = 43,366.275$, and the orbital period is estimated as $P = 26.4960$ days (e.g., see \cite{gregory2002}). In quoting orbital phases, we have assumed that the orbital phase of periastron is $\phi_{\rm peri} = 0.23$\cite{aragona2009} as is common in the study of this source (but see {\bf Methods} for further discussion of orbital uncertainties). 
%


\begin{table}
\scriptsize
\caption{Log of observations}
\vspace{-1cm}
\label{tab:obs}
\medskip
\begin{center}
\begin{tabular}{c c c c c c}
\hline \hline
Mid of observation time & Orbital phase & Exposure Time/h & Sampling Time/$\mu$s & Pulse detected & S$_{\rm mean}$/S$_{\rm UL}$/$\mu$Jy\\
\hline
58,788.7257  & 0.07  & 2.2 &  98.304 & No  & --/1.61 \\
58,855.5278  & 0.59  & 3.0 &  98.304 & Yes & 4.40/1.37 \\
59,093.8646  & 0.58  & 3.0 & 196.608 & No  & --/1.37 \\
59,094.8681  & 0.62  & 2.0 & 196.608 & No  & --/1.68 \\
\hline
\specialrule{0.05em}{2pt}{2pt}
\end{tabular}
\\
\setlength{\parskip}{-1.0ex}
\item The orbital phases are calculated with the radio ephemeris given in Ref.\cite{gregory2002}.
\end{center}
\end{table}

The PRESTO (PulsaR Exploration and Search TOolkit package\cite{Ransom2002} was used to search for the periodic signal. The Dispersion Measure (DM) in the direction of \lsi{}, as predicted by the YMW16 model\cite{yao2017} is 138.58 pc cm$^{-3}$, while its upper limit is 316.59 pc cm$^{-3}$. To avoid missing the signal, and considering the error, the DM range for our search was 0 to 500 pc cm$^{-3}$. Since the orbital period is much longer than the exposure time, the Doppler effect is negligible. We have carried out the acceleration search with {zmax} $= 200$, we also used the routine BEAR for single pulse search (see details in {\bf Methods}). 

\begin{figure*}[htbp]
\centering
\includegraphics{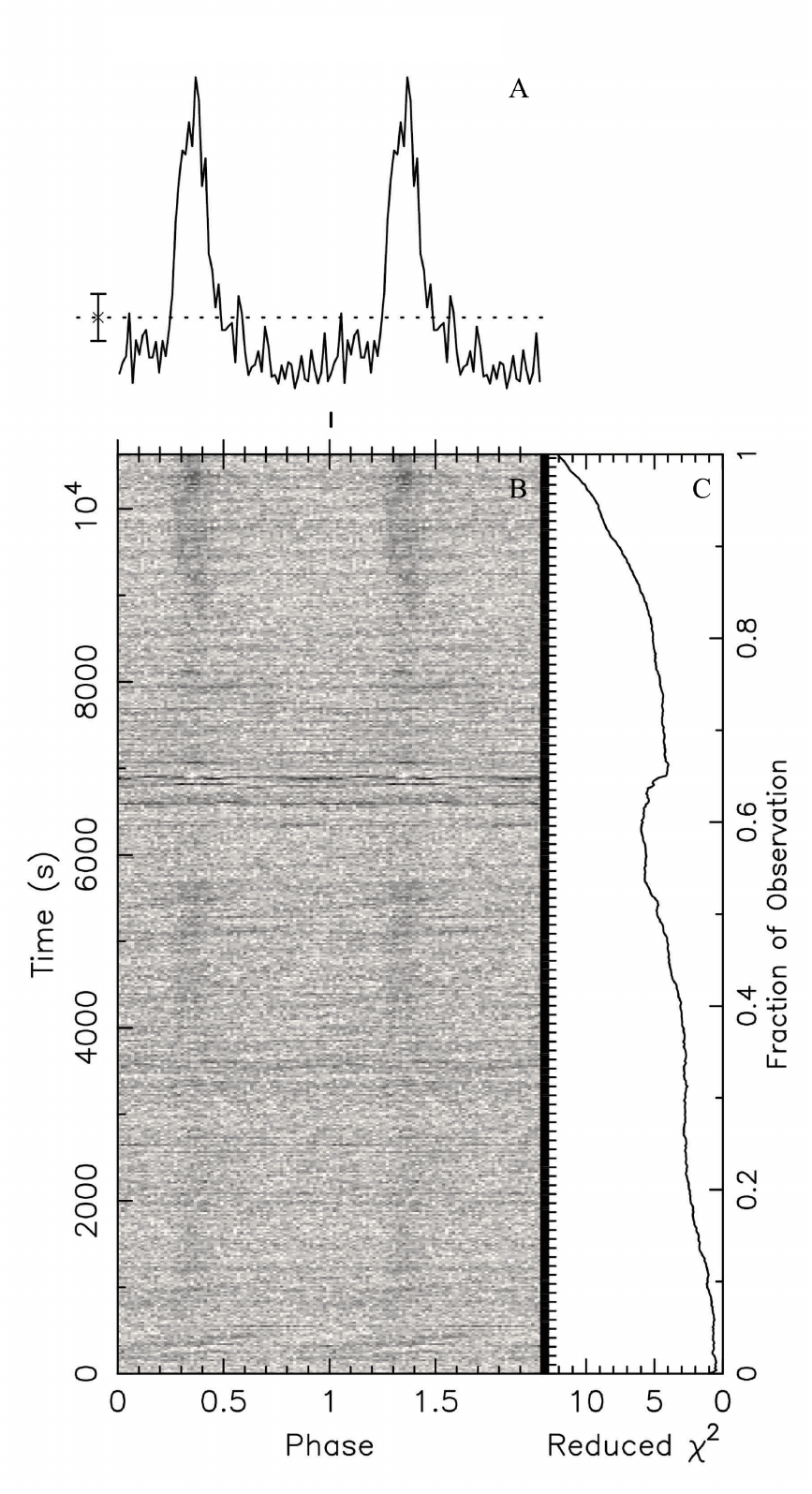}
\caption{{\bf Radio pulsations were recorded in the \fast{} data recorded on 2020 January 7$^{th}$.}  The data are folded over the period of $P = 269.15508$ ms, and the reduced $\chi^{2}$ (i.e. the significance of the pulsation) is computed with the folded profile compared to the average value of the folded profile. The integrated pulse profile is plotted in A. The intensity (gray-scale) of the pulse and the reduced $\chi^{2}$ as a function of the pulse phase and the observational time are shown in B and C, respectively. \label{fig:pulse}}
\end{figure*}

An unambiguous pulse signal (a single-trial significance of $\sim$22.4$\sigma$, see {\bf Methods}) with a single-peak profile emerges from the data taken on 2020 January 7$^{th}$ (MJD = 58,855.5278, Figure \ref{fig:pulse}). The period, pulse width and DM of this pulsar are 269.15508(16) ms, $33.30\pm0.96$ ms, and 240.1 pc cm$^{-3}$, respectively (see also Supplementary Figure 1). The pulsations disappeared in the 3rd and 4th observations (one-day apart of each other), taken several months after the positive detection, at a similar orbital phase. A single pulse search was conducted for our observation, and more than 40 were detected in the second observation (where the radio pulsation is visible), but none were seen in the other three (see {\bf Methods}).

Given that our observations are short in comparison to the orbital period of the binary, and the pulsation appears to be non-steady in nature, the orbital imprint cannot be detected in our data. In addition, the angular resolution (L-band) of \fast{} is of $\sim 2.9^{'}$. As for other \fast\ observations in the same band, we cannot therefore formally exclude the presence of a pulsar just behind LS I +61° 303, unrelated to it, that is responsible for the pulsed emission. This is deemed unlikely, however ({\bf Methods}).


The fact that the pulsations are not present in three out of four observations (a couple of them very close to one another) reveals a rapid change of conditions either in the interstellar medium between us and the source,
the neutron star environment, or the neutron star itself.

%
%
Interstellar scintillation may be produced by a diffractive scattering medium along our line of sight, 
for which the typical timescale\cite{Cordes1998} is $\Delta t_d = 2.53 \times 10^4 (D \Delta \nu _d)^{1/2}/(\nu V_{\rm ISS})$ s, where $D$ the distance to the source in kpc, $\Delta \nu _d$ the decorrelation bandwidth in MHz, $\nu$ the observed frequency in GHz and $V_{\rm ISS}$ the velocity of interstellar scintillation (ISS) diffraction pattern in km\,s$^{-1}$ (which in binaries is typically dominated by pulsar velocities, in the range of few tens to few hundreds km\,s$^{-1}$). For the observation shown in Figure 1, scintillation could explain changes in flux up to a few minutes, but not the several hours, in consecutive days, when pulsations are absent. And this does not consider that the bandwidth of the observation is larger than the decorrelation bandwidth and therefore scintillation would be averaged out. Refractive scintillation, on the other hand, could have a longer timescale and we cannot rule it out a priori, although  the flux modulation we have seen would be larger than expected\cite{Narayan1992}.

%
A second possibility is that the flux modulation is related to an intrinsic nulling of the pulsar, a temporary broadband cessation of normal pulsar emission likely produced by magnetospheric changes. About 8\% of the known pulsars exhibit nulling\cite{Sheikh2021}, with durations that go from a single pulse to several days. Some pulsars actually live most of the time in their null state, to activate themselves only for short periods, making them difficult to detect. Mode changing is a related phenomenon where the mean pulse profile abruptly changes in a matter of minutes to come back to the original later on (see, e.g.,\cite{Wang2007}). With the current lack of information regarding pulsations at other frequencies, associating the absence of pulsations we observe to nulling or mode changing is not possible. The fairly rapid spin period, and the plausibly low age of \lsi\ (see e.g.,\cite{dubus2013}), may argue against these interpretations. Nulling appears predominantly for longer-period and older pulsars. It is interesting, in the same context, to note the similarity of the phenomenology seen to those of rotating radio transients, pulsars which sporadically emit single pulse outbursts instead of continuous pulse trains\cite{McLaughlin2006}.

Finally, changes in the wind properties can easily affect the pulsed signal. And as the Be stellar wind is likely to be clumpy,  it would be impossible to predict the absorption level that any radio signal will be subject to in a local basis (see the discussion in e.g.\cite{Zdziarski2010}). Then, the transient behaviour could plausibly be interpreted as a result of the rapid change in the environmental conditions. Regarding this possibility, it might be useful to compare with the case of the pulsar-composed gamma-ray binary PSR B1259-63. Despite this system has a much larger orbit, radio flux variations at a time scale of minutes to hours were also reported in PSR~B1259-63, together with changes  in the local properties of the Be star wind/disk encountered by the pulsar (see e.g., \cite{Johnston2005} and references therein). It is reasonable to expect these same effects apply to \lsi, enhanced due to the smaller spatial scale of the system.  

We can also draw some phenomenological connections with transitional pulsars. These are redback systems, a millisecond pulsar in a close orbit with a low-mass companion star. These systems can exhibit highly variable eclipses from orbit to orbit, radio variability, X-ray/radio anti-correlation, and some also fully disappear in radio for days (see, e.g., \cite{Bogdanov2018}). Whereas \lsi\ is not a redback pulsar, similarly to them, it does have a non-compact stellar companion which fills the system with ionized gas. The variability of the pulsations could likely be due to the latter, and the pulsed signal can also be affected by matter-magnetosphere interaction. Future simultaneous campaigns  in X-rays and radio may shed further light on this comparison.


In any case, our finding suggests that  \lsi\ is a  pulsar-composed system and place the tight constraint on pulsar-based models when explaining its multifrequency phenomenology.  Among them, models based in state transitions might be appealing: A study in the context of  the short magnetar-like flares and the superorbital variability  allowed to define a preferred pulsar period for \lsi\ of 0.26 s (see figure 4 of Ref.\cite{Papitto2012}), as we find here. With the knowledge gained from our observational finding, this and other pulsar- based models can be revisited and enhanced, and applied to the search and characterization of pulsations at all frequencies.





\clearpage

\begin{methods}

\subsection{\fast{} observations}~

\fast\/ is the most sensitive operating radio telescope\cite{Nan2011,jiang2019}, and it has recently been used to detect some very faint pulsars\cite{han2021,Qian2020,Pan2021}. \lsi{} was observed by \fast{}  four times in 2019--2020, and all data were equally processed with the packages available in PRESTO using the following usual steps\cite{Ransom2002}: 1. We masked and zapped the radio-frequency interference (RFI) using the routine \texttt{rfifind}; 2. After RFI excision, the data  were de-dispered with the trial DMs between 0 and 500 pc~cm$^{-3}$ by using the routine \texttt{prepsubband}; 3. For the resulting de-dispered time series, we carried out a blind Fourier-domain acceleration periodicity search with the routine \texttt{realfft} and \texttt{accelsearch}, yielding the periodic candidates. 4. The periodic candidates were further sifted with the routine \texttt{ACCEL$\_$sift.py}. 5. The data were corrected to Solar System Barycentre and were automatically folded over all derived periodic candidates and possible period-derivatives using the routine \texttt{prepfold}. We could then check and confirm the signal by diagnosing the folding plots. That is, the precise values of DM, $P$, $\dot{P}$ and  their uncertainties were obtained by folding the data to reach a maximum $\chi^{2}$ (i.e. SNR).

The pulsations were only detected in the \fast{} data taken on 2021 January 7$^{th}$, and the folding results with the identified period, $P=269.15508 \pm 0.00016$ ms\footnote{We made the barycenter correction with the wrong coordinates, and the results given in the Atel \#14297 were inaccurate. The results are corrected in this paper.}, are shown in Supplementary Figure 1. We did not find any public pointed observation around 2020 January 7$^{th}$ in multi-wavelength bands. We checked that there was no unusual behavior displayed in survey data, including Swift/BAT, MAXI, and {\it Fermi}-LAT.

Due to the fact that we are considering only 3 hours of observation spanning a very small orbital phase, we cannot recover the Doppler-shifted signals to determine this pulsar's intrinsic $\dot{P}$ and $\ddot{P}$. When carrying out a non-acceleration search with the option ``-nopdsearchr" in the routine  \texttt{prepfold}, we obtain a slightly smaller single-trial significance of $\sim 21.1\sigma$.

An estimate of the mean flux density that each of our observations would have detected, $S_{\rm mean}$, can be obtained via\cite{Lynch2011}:
\begin{equation}
S_{\rm mean}=\frac{\eta \beta T_{\rm sys}}{G\sqrt{N_{\rm p}\Delta \nu T_{\rm int}}}\sqrt{\frac{W}{P-W}}
\end{equation}
where $\eta$ is the SNR threshold, $\beta$ is the sampling efficiency, $T_{\rm sys}$ is the system temperature, $T_{\rm int}$ is the integration time, and $G$ represents for antenna gain. For \fast{}, $\beta$=1, $T_{\rm sys}$=24 K, $G$=16 K Jy$^{-1}$. The bandwidth $\Delta\nu$ is 300 MHz as we masked 100 MHz in our pipeline due to the RFI, and finally, $N_{\rm p}$ = 2 is the number of polarizations, $P$ is the period of pulsar, and $W$ is the width of pulse profile. The value of $W/P$=0.1 (see the figure in the main text) is adopted for \lsi{}. The 7$\sigma$ detection limit ($\alpha$=7) of the flux density for each observation, $S_{\rm UL}$, is also calculated and listed in Table \ref{tab:obs}.  

In comparison, the Green Bank Telescope (GBT, {\protect\url{https://greenbankobservatory.org/science/telescopes/gbt/}}) and the 100-m Effelsberg radio telescope ({\protect\url{https://www.mpifr-bonn.mpg.de/effelsberg/astronomers}}) have similar system temperatures, but lower antenna gains,  2 K Jy$^{-1}$  and 1.55 K Jy$^{-1}$ at 1.4 GHz, respectively. That is, the mean pulsed signal reported with \fast{} data ($\sim$4.4 $\mu$Jy, see Table \ref{tab:obs}) would be unattainable for the GBT and Effelsberg telescopes: an unrealistic integration time of $> 20-30$ hours would have been required for a 7$\sigma$ detection. Alternatively, flux variations, as are evidently displayed in \lsi\ may lead to overpassing the threshold for detection at certain times. According to Equation 1, the estimated pulsed flux of \lsi{} increased up to 12.85 $\mu$Jy (single-trial significance of $26.7 \sigma$) in the last half-hour. Such level of flux might be detected by the GBT and Effelsberg telescopes within a reasonable observational time of $\sim$2.2 hours. \lsi{} is too north to be observed by the Arecibo ({\protect\url{http://www.naic.edu/science/generalinfo_set.htm}}).

\subsection{Single pulse analysis}~

In addition to PRESTO, we have also used the package BEAR (Burst Emission Automatic Roger)\cite{BEAR} to do a single pulse search for all observational data. The RFI of \fast{} are mostly coming from satellites, so the frequency from 1200 MHz to 1240 MHz and 1270 MHz to 1300 MHz were cut off from the data. We de-dispersed the data from 0 pc cm$^{-3}$ to 500 pc cm$^{-3}$ in steps of 0.5 pc cm$^{-3}$ and use a box-car-shaped match filter to search for bursts with width between 0.2 ms and 30 ms. Candidates with a SNR threshold larger than 7 were plotted by BEAR and visually inspected. For our data, the observations performed on MJD 58,788, 59,093 and 59,094 lead to no significant detection in our single pulse searching pipeline. The analysis of the data obtained on MJD 58855 leads to 42 single pulses. Supplementary Figure 2 shows all of the single pulses' dynamic spectra and their profiles. Parameters of all these single pulses are listed in Supplementary Table 1. The DM value for each single pulse (DM$_{s/n}$) was obtained by aligning the signal across frequency to achieve the best peak SNR\cite{LK12HPA}. This also explains the apparent variability seen in the DM values, meanwhile the mean value of DM is 240.2 pc cm$^{-3}$ which is consistent with the pulsar mentioned above. We use intensity weighted width (IWW) to measure the burst width by treating the pulse profile as the temporal intensity distribution function, and then calculated the standard deviation of time. The mean flux density is computed using the following equation, uncertainties are dominated by the system noise temperature ($\sim20$\% \cite{jiang2019}).
\begin{equation}
S_{\rm mean}=\frac{\text{SNR} \cdot T_{\rm sys}}{G\sqrt{N_{\rm p}T_{\rm sample}\Delta \nu}}
\end{equation}

We show the folded single pulses data with ${P}$ and $\dot{P}$ obtained in Supplementary Figure 1, and display their occurrences in the time-phase diagram with the red bars (Supplementary Figure 3).

The non-zero DM value and the variable RFI recorded in the data suggested that the detected pulses are unlikely from the instrumental or terrestrial interference. Plus, the single pulses displayed a dramatic flux variation (by a factor of $> 10^{3}$, see Supplementary Table 1) in short time scales, while retaining the DM value. Therefore, the detections of both the weak averaged pulse and the energetic single pulses in the same observation cannot be interpreted as instrumental origin or terrestrial interference. 

In addition, the detection of energetic single pulses would indicate that the emission is unlikely from the secondary lobes, i.e., emission from a pulsar far away the field of \lsi{} ($\sim 6^{'}-8^{'}$\cite{Jiang2020}).  If the emission would come from secondary lobes, the intrinsic flux density should be about $\sim 1000$ times brighter than the detected level\cite{Jiang2020}, i.e. about $\sim 10-100$ Jy. If so, other telescopes could have easily detected the signal already. While this paper was being reviewed, we made another dozen \fast{} observations, covering the whole orbital phase. The preliminary analysis reveals several single pulses on 2021 November 2nd, corresponding to an orbital phase of $\sim 0.69$. These single pulses share similar properties to those reported here in more detail and further support their origin in \lsi{}. A detailed analysis on these single pulses will be presented elsewhere.

Pulse No.24 and Pulse No.41 show an exponential-like scattering tail. For these two bursts, we used a Gaussian convolved with a one-sided exponential function to fit them,
\begin{equation}
f(t,\tau)=\frac{S}{2\tau}\text{exp}\left({\frac{\sigma ^{2}}{2\tau^{2}}}\right)\text{exp}\left({-\frac{t-\mu }{\tau }}\right)\times\left\{ 1+\text{erf}\left[\frac{t-(\mu+\sigma^{2}/\tau)}{\sigma\sqrt{2}}\right]\right\}
\end{equation}
where $S$ is the flux density of the Gaussian, $\mu$ is its center, and $\sigma$ is its standard deviation. $\tau$ is a time constant of the one-sided exponential function. We split the data into 8 evenly spaced sub-bands across the 500\,MHz raw bandwidth, then clip the channels RFI. For each sub-band with SNR $\ge$ 7, we integrated the pulse intensities over time and used MCMC to fit the intensity profile with the equation above along the frequency axis (panel B of Supplementary Figures 4 and 5) in order to get the scattering time scale ($\sigma_{\rm chn}$) and the standard deviation ($\tau_{\rm chn}$) of each sub-band (panel  C of Supplementary Figures 4 and 5).  At each frequency channel, the scattering time scale is
\begin{equation}
\tau_{\rm chn}(\nu)=\tau_{\rm chn}\left(\frac{\nu}{\nu_{\text{ref}}}\right)^{\alpha}
\end{equation}
where $\nu_{\rm ref}$ is the reference frequency and is set to 1 GHz, $\alpha$ is the frequency scaling index. Linear regression of the scattering timescales shows their scattering timescale at 1 GHz of $\tau_{\rm 1GHz}$=29.85$\pm$4.64 ms, 15.57$\pm$2.39 ms and a frequency scaling index of $\alpha$=-2.49$\pm$0.53, and $-1.91\pm0.80$  for the two pulses, respectively. The scattering by the ionized plasma leads to the pulse being asymmetrically broader at lower frequencies. For the thin-screen scattering model, we could expect scaling indexes of -4 and -4.4, for Gaussian and Kolmogorov inhomogeneities, respectively\cite{Lang1971, Romani1986}. However, note that deviations from the theoretical models had been already reported in several pulsars (e.g.  PSR B0823+26, PSR B1839+56, and others): It was suggested that lower $\alpha$ values could result from limitations of the thin-screen model or from an anisotropic scattering mechanism\cite{Bansal2019}. In our case, the distribution of the ionized plasma around \lsi{}  cannot be an infinite thin screen, and should deviate from either Gaussian or Kolmogorov inhomogeneous distribution.



\subsection{Uncertainties}~

Except for the orbital period, the other orbital parameters, including the  orbital phase of periastron and the eccentricity, still bear relatively large uncertainties. Currently, the orbital period is well-determined as $P \sim 26.4960\pm0.0028$ days, see e.g.\cite{gregory2002}. The orbital phase of periastron ($\phi_ {\rm peri} $), instead, was estimated by fitting the radial velocities to be in the range of 0.23--0.30, and the eccentricity  $e$ in the range of 0.55--0.72, e.g.,\cite{casares2005,grundstrom2007,aragona2009}. Recently, however, Kravtsov et al.\cite{Kravtsov2020} obtained a notably different orbital solution using optical polarization measurements, i.e.  a small eccentricity $e < 0.2$, and $\phi_ {\rm peri} \sim 0.6$, although some of their parameters are degenerate. The latter solution, as the authors discuss, is not devoid assumptions that may directly impact on the results. If correct, the solution presented by Kravtsov et al. would imply that notable reassessments are needed when considering, for instance, the orbital location and interpretation of all multifrequency phenomenology of the system, as well its age. However, we advise to keep this uncertainty in mind until a final orbital solution is established beyond doubt. Given that our observations are short in comparison to the orbital period of the binary, and that the pulsation appears to be non-steady in nature, the orbital imprint cannot be constrained with our data.
%


Finally, we ponder how likely could it be that within this spatial extent, \fast\ is actually detecting a different pulsar lying within the angular resolution (at the L-band) of \fast{}, $\sim 2.9^{'}$\cite{jiang2019}. For \lsi\ in particular, a similar issue appeared when analyzing the magnetar-like flares detected from the same region\cite{depasquale2008,barthelmy2008,Arjonilla2009}. Swift, the X-ray satellite that observed them, had a $\sim 1.4^{'}$ positional uncertainty and no other candidate different from \lsi\ was found. A reanalysis of the flare data, as well as a subsequent 96 ks observations with the {\it Chandra} X-ray telescope,  did not reveal any other candidate in that region either\cite{Torres2012}. Neither it did a combined analysis of archive Very Large Array radio data nor near infrared observations\cite{Arjonilla2009}. These studies concluded that the simplest explanation is that \lsi\ was the origin of the flares.

There are just a few gamma-ray binaries known in the Galaxy (less than 10 in total) and we know already about three thousand radio pulsars\cite{manchester2005}, of which about 30 have shown magnetar-flare behavior. If we were to assume that \lsi\ is different from the radio pulsar we detect, and different too from the origin of the magnetar-like flares detected from the same region earlier, we would need to find three relatively rare objects aligned within a few arcmin. To qualitatively assess how likely this is we can consider first  that  the short bursts and the radio pulsations reported here come from two unrelated neutron stars in the small field of view close to \lsi{}. Using the ATNF Pulsar Catalogue ({\protect\url{https://www.atnf.csiro.au/research/pulsar/psrcat/}}) as a basis for what we have been able to detect with current instruments,  taking all 2072 pulsars within the Galactic latitude of $\pm$10$^o$  and excluding those in globular clusters, the probability to find two pulsars within $\sim 2.9^{'}$ is $<7\times10^{-6}$.  We would still need to multiply this by the probability to find \lsi\ within the same region. Assuming that there are 10 sources like \lsi, the probability would reduce to $<1\times10^{-11}$. If we were to assume that the system producing the magnetar-like flare and the pulsation we detect are the same, i.e., a single pulsar, but different from \lsi, the probability to randomly find both a pulsar producing magnetar phenomenology and a gamma-ray binary within  $\sim 2.9^{'}$ assuming a uniform distribution in the $\pm$10$^o$ Galactic plane region, would be $\sim 3 \times 10^{-10}$. 

We caveat that these numbers are uncertain, as they lack details regarding spatial distribution of sources being non-uniform or considerations relative to the age the system. The magnetar-behavior, for instance, seems to appear more often at younger pulsar ages, so not all pulsars would equally serve as a counterpart for the flares, see e.g.,\cite{Kaspi2017}. These estimations also suffer from biases from incomplete and non-uniform observational samples. The ATNF catalog is a multi-frequency, multi-facility compilation, rather than a complete survey at a fixed sensitivity. Ideally, one would use a complete survey using \fast\ itself to judge the surface density of pulsars specifically at the detected flux density and band. Although this is currently unavailable, simulations\cite{han2021} showed that \fast\ should  able to discover about 1000 pulsars, depending on available observation time. This number is lower than what we have considered above using the whole ATNF catalog and does not change the prior conclusion.

Considering the reverse the problem can be useful: we asked how many pulsars with magnetar behaviour there should be for an alignment with \lsi\ to happen by chance. To do this we can simulate sets of Galactic positions of putative pulsars producing magnetar flares  and measure which is the random coincidence between these and our source of interest (so that both sources lie within 3 arcmin). We can do so respecting the spatial distribution of the current population of pulsars in Galactic longitude and latitude. Using 100000 simulated sets (a larger number does not notably change the results), of 1928 magnetars each, the average number of simulated coincidences between the position of one of them and \lsi\ would be $\sim 0.00093$ (standard deviation 0.0304). In this example, we have taken the actual number of known pulsars with $|b|<5$ degrees (1928) and considered that future samples will contain such a number of magnetars. This is indeed conservative. For context, this number is a factor of $>60$ beyond the magnetars currently known, or a factor of 2 larger than all pulsars expected to be detected anew by \fast, or even a factor of 4 larger than the number of magnetars  born in the Galaxy in the last 25 kyr assuming the most favorable birth rate\cite{Beniamini2019}. The main conclusion is that although it is formally impossible to rule out that there is a projected superposition of different sources, the combinations of these relatively rare systems in such a small region of the sky appears to be unlikely. 


\begin{addendum}

\item[Code availability] ~

PRESTO (\protect\url{https://www.cv.nrao.edu/~sransom/presto/})

BEAR (\protect\url{https://psr.pku.edu.cn/index.php/publications/software/})

\item[Data Availability] The data sets generated during and/or analysed during the current study are available from the authors on reasonable request.

\end{addendum}
\end{methods}

\begin{addendum}
 \item This work made use of the data from \fast{} (Five-hundred-meter Aperture Spherical radio Telescope). \fast{} is a Chinese national mega-science facility, operated by National Astronomical Observatories, Chinese Academy of Sciences. We acknowledge the use of the ATNF Pulsar Catalogue.  S.S.W. and B.J.W. thank Dr. Zhichen Pan for discussions on the \fast{} data analysis. S.S.W. thank Profs. Zhong-Xiang Wang, Shuang-Nan Zhang and Kejia Lee for many valuable discussions.  J.L., D.F.T. and A.P. acknowledge discussions with the international team on `Understanding and unifying the gamma-rays emitting scenarios in high mass and low mass X-ray binaries' of the ISSI (International Space Science Institute), Beijing. The authors thank the support from the National Key R\&D program of China No. 2017YFA0402602, 2021YFA0718500, National SKA Program of China No. 2020SKA0120100, No. 2020SKA0120201, the National Natural Science Foundation of China under Grants U2038103, 11733009, U2031205, U1938109,  11873032, the Youth Innovation Promotion Association of CAS (id. 2018075), the Chinese Academy of Sciences Presidential Fellowship Initiative 2021VMA0001, the National Foreign Experts Program of Ministry of Science and Technology BB504000808, and the international visiting professorship program of  the University of Science and Technology of China. S.S.W. acknowledges the financial support by the Jiangsu Qing Lan Project.  D.F.T. also acknowledges grants PID2021-124581OB-I00, PGC2018-095512-B-I00 as well as the Spanish program Unidad de Excelencia ``María de Maeztu'' CEX2020-001058-M.  A.P. acknowledges financial support from the Italian Space Agency (ASI) and National Institute for Astrophysics (INAF) under agreements ASI-INAF I/037/12/0 and ASI-INAF n.2017-14-H.0, from INAF 'Sostegno alla ricerca scientifica main streams dell'INAF', Presidential Decree 43/2018, and from PHAROS COST Action N. 16214.

 \item[Author contributions] S.S.W., L.Q., and B.J.W. have contributed equally to these results. S.S.W. proposed the observational project. The \fast{} team leaded by P.J. designed and scheduled the observations during the \fast{} commissioning stage. L.Q. carried out the  observations, and B.J.W. analyzed the data. D.F.T., J.L. and A.P. contributed to interpreting the results. D.F.T., S.S.W., B.J.W. wrote the paper. P.J., R.X.X., J.Z.Y., Q.Z.L., M.Y.G., Q.R.Y. participated in the interpretation of the results. All authors discussed the contents of the paper and contributed to the preparation of the manuscript.

 \item[Competing Interests] The authors declare that they have no competing financial interests.

 \item[Correspondence and request for materials] Correspondence and requests for materials should be addressed to S.S.W.~(email: wengss@njnu.edu.cn) and D. F. Torres~(email: dtorres@ice.csic.es).

\end{addendum}


\section*{Supplementary Information}

\renewcommand{\figurename}{\textbf{Supplementary Figure}}
\renewcommand{\tablename}{\textbf{Supplementary Table}}
\setcounter{figure}{0}    

\begin{figure*}
\hspace{-1cm}
\includegraphics[width=0.85\textwidth,angle=90]{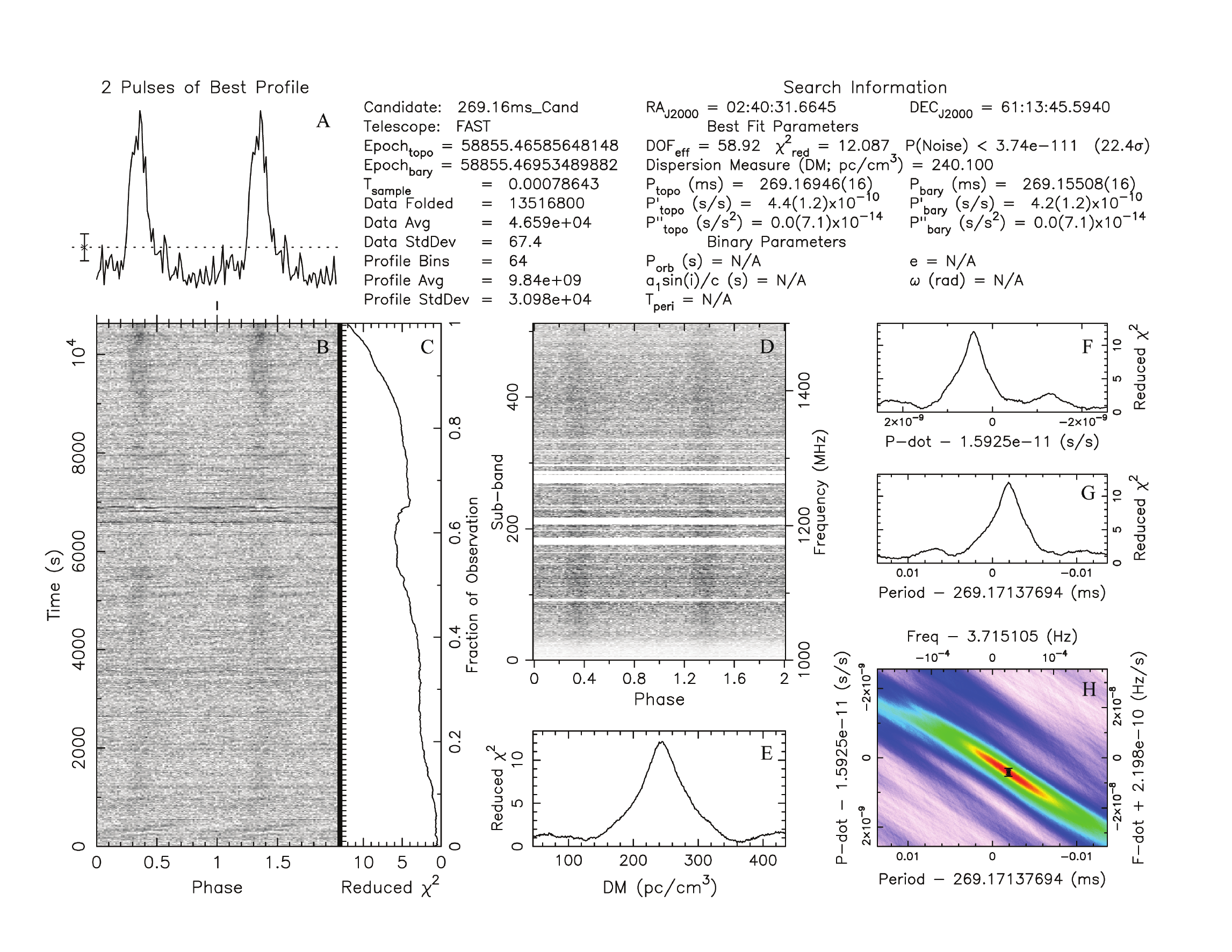}
\caption{{Standard plot of the radio-pulsation search results with the routine \texttt{prepfold} in the PRESTO package.} The  validity of derived parameters can be found in the panels {\bf E-G}, where the $\chi^{2}$ is a function of DM, $P$, and $\dot{P}$, respectively. Meanwhile, the confidence contour of $P$ and $\dot{P}$ is shown in {\bf H}. The averaged pulse profile as a function of observing frequency is shown {\bf D}.  Pulse profiles in the left-hand plots are magnified and shown in Figure 1 of the main text.  \label{fig:lsi_pulse}}
\end{figure*}

\begin{figure*}
    \centering
    \includegraphics[width=0.9\textwidth]{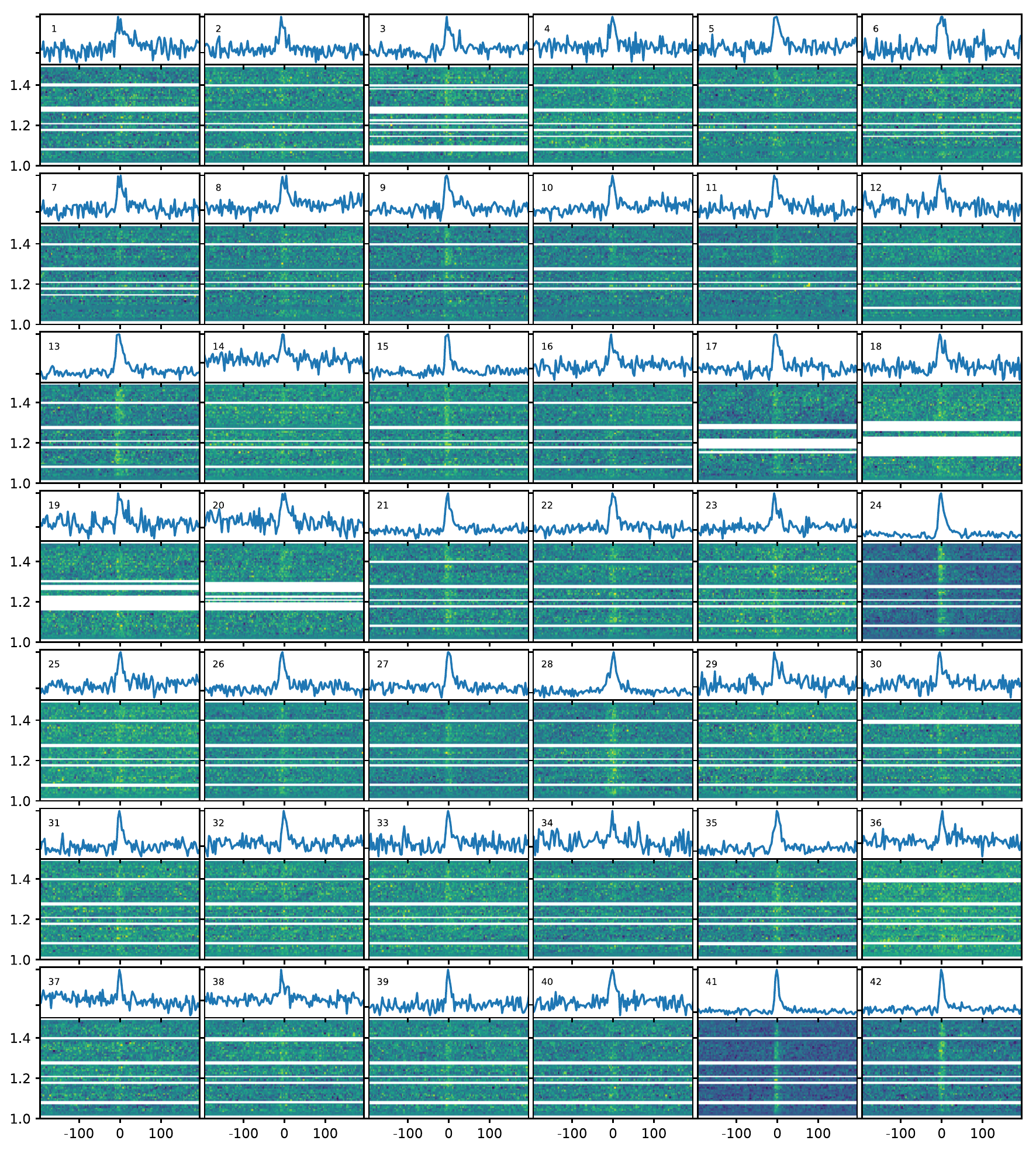}
    \caption{{Single pulses detected in the \fast{} data for 2020 January 7$^{th}$.} The pulse profile and the dynamical spectrum are shown for each single pulse. The burst number corresponding to the burst number in Extended Data Table \ref{tab:frb} are given in each panel. White strips in the dynamical spectra indicate the RFI zapping. The profile of the single pulse varies from each other, as is commonly found in other pulsars.}
    \label{fig:lsiallsp}
\end{figure*}

\begin{figure*}
    \centering
    \includegraphics{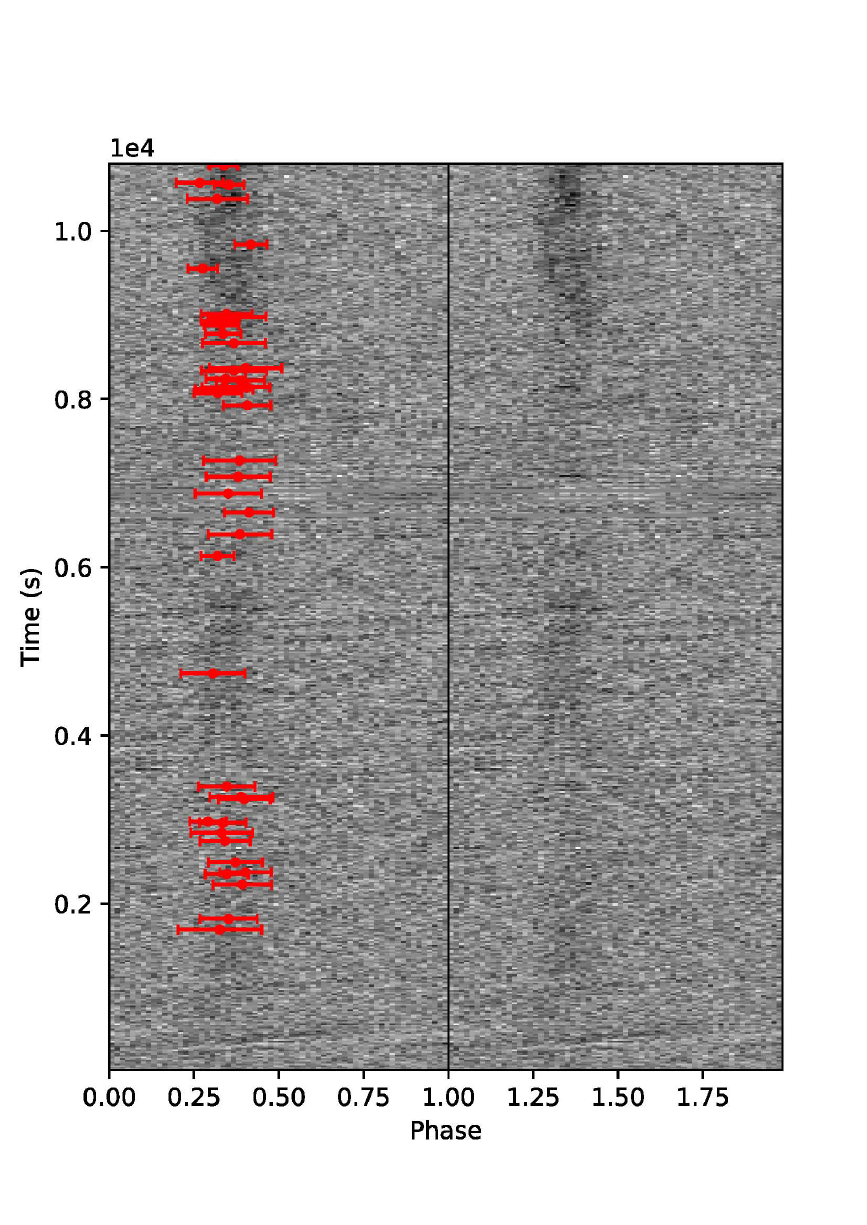}
    \caption{{The intensity versus the pulse phase and the observational time.} The red bars mark the pulse phase and the occurrence time of the single pulses having a S/N ratio larger than 7.}
    \label{fig:sp_tvsphase}
\end{figure*}

\begin{figure*}
    \centering
    \includegraphics{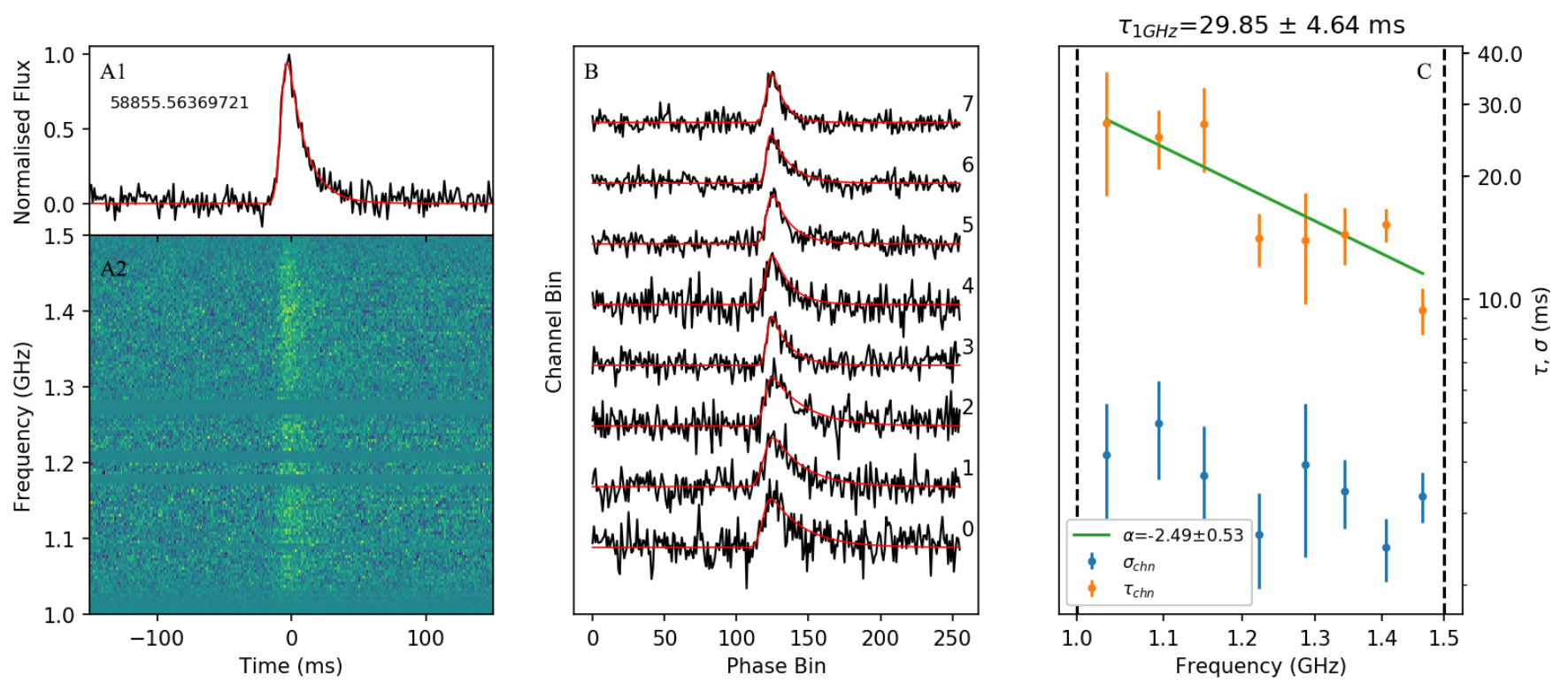}
    \caption{{ Pulse No. 24.} {\bf A1} and {\bf A2} show the  pulse profile  and dynamical spectra, respectively. {\bf B}: fitting of sub-band profile which with $S/N \ge 7$. {\bf C}: scattering timescales as a function of frequency. The fitting parameters $\alpha$, $\sigma_{chn}$, $\tau_{chn}$ and their 1$\sigma$ errors are plotted.}
    \label{fig:scat_1}
\end{figure*}

\begin{figure*}
    \centering
    \includegraphics{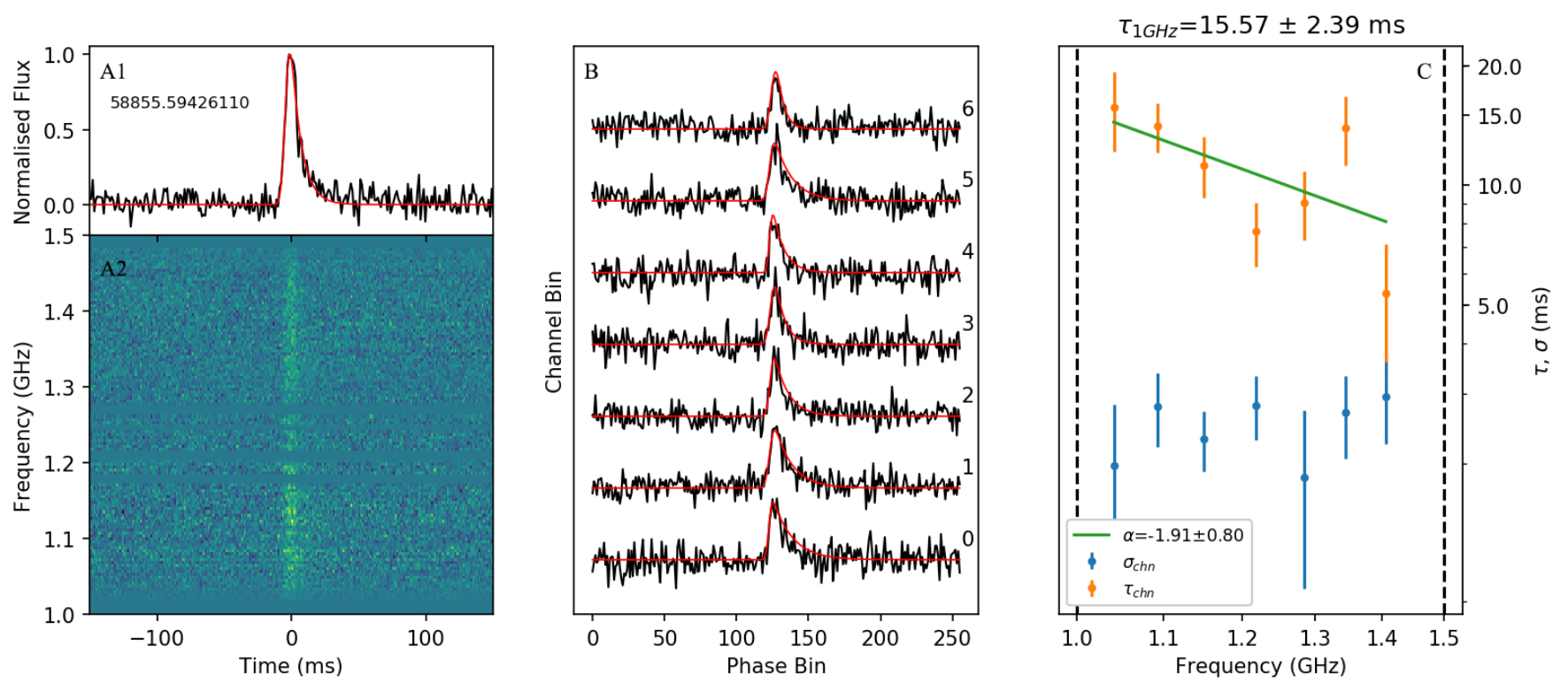}
    \caption{{Pulse No. 41.}  {\bf A1} and {\bf A2} show the  pulse profile  and dynamical spectra, respectively. {\bf B}: fitting of sub-band profile which with $S/N \ge 7$. {\bf C}: scattering timescales as a function of frequency. The fitting parameters $\alpha$, $\sigma_{chn}$, $\tau_{chn}$ and their 1$\sigma$ errors are plotted.}
    \label{fig:scat_2}
\end{figure*}

\begin{table}
\caption{Parameters of single pulses}
\footnotesize
\vspace{-1cm}
\scriptsize
\label{tab:frb}
\medskip
\begin{center}
\begin{tabular}{c c c c c c }
\hline 
Burst No. & Barycentric TOA\ & DM$_{s/n}$/pc cm$^{-3}$ & Pulse Width/ms & S$_{\rm mean}$/mJy & SNR \\
\hline

 1 &  58855.48912132 & 239.7$\pm$6.0 & 33.30$\pm$2.15 & 4  & 12 \\
 2 &  58855.49060114 & 239.5$\pm$4.1 & 22.79$\pm$2.70 & 5  & 12 \\
 3 &  58855.49533330 & 239.8$\pm$3.3 & 23.18$\pm$2.29 & 6  & 15 \\
 4 &  58855.49677862 & 243.0$\pm$3.4 & 17.07$\pm$2.13 & 5  & 11 \\
 5 &  58855.49701244 & 240.4$\pm$3.0 & 20.55$\pm$2.03 & 6  & 15 \\
 6 &  58855.49838616 & 236.9$\pm$4.2 & 21.45$\pm$2.80 & 5  & 11 \\
 7 &  58855.50132061 & 243.1$\pm$3.6 & 19.95$\pm$2.21 & 5  & 12 \\
 8 &  58855.50245141 & 242.5$\pm$3.8 & 24.58$\pm$2.57 & 5  & 14 \\
 9 &  58855.50381590 & 242.0$\pm$2.5 & 18.57$\pm$1.59 & 7  & 16 \\
10 &  58855.50399956 & 240.3$\pm$2.4 & 14.48$\pm$1.57 & 7  & 13 \\
11 &  58855.50708709 & 245.9$\pm$3.2 & 20.70$\pm$2.09 & 6  & 14 \\
12 &  58855.50738301 & 235.2$\pm$5.4 & 25.07$\pm$3.64 & 4  & 10 \\
13 &  58855.50881900 & 242.5$\pm$1.9 & 22.48$\pm$1.27 & 11 & 26 \\
14 &  58855.52436392 & 243.5$\pm$6.1 & 25.48$\pm$4.27 & 4  & 9  \\
15 &  58855.54052275 & 238.1$\pm$1.3 & 13.12$\pm$0.86 & 12 & 22 \\
16 &  58855.54351360 & 237.8$\pm$4.6 & 25.29$\pm$2.97 & 5  & 12 \\
17 &  58855.54650744 & 238.6$\pm$2.3 & 19.32$\pm$1.53 & 8  & 18 \\
18 &  58855.54912095 & 241.7$\pm$4.1 & 26.37$\pm$2.72 & 5  & 14 \\
19 &  58855.55143568 & 237.9$\pm$5.0 & 25.29$\pm$3.28 & 4  & 11 \\
20 &  58855.55369114 & 240.0$\pm$5.6 & 28.58$\pm$3.60 & 4  & 11 \\
21 &  58855.56126751 & 237.4$\pm$1.8 & 18.69$\pm$1.21 & 10 & 22 \\
22 &  58855.56294326 & 242.0$\pm$2.3 & 19.02$\pm$1.42 & 8  & 18 \\
23 &  58855.56346356 & 240.1$\pm$2.9 & 22.91$\pm$1.89 & 7  & 17 \\
24 &  58855.56369721 & 240.7$\pm$1.9 & 37.82$\pm$2.03 & 14 & 44 \\
25 &  58855.56375349 & 241.3$\pm$2.8 & 18.31$\pm$1.79 & 7  & 14 \\
26 &  58855.56468178 & 240.0$\pm$2.3 & 18.99$\pm$1.52 & 8  & 18 \\
27 &  58855.56491529 & 239.2$\pm$1.8 & 15.67$\pm$1.16 & 10 & 19 \\
28 &  58855.56603998 & 238.5$\pm$1.9 & 25.93$\pm$1.25 & 11 & 29 \\
29 &  58855.56637341 & 237.8$\pm$5.2 & 28.66$\pm$3.18 & 4  & 12 \\
30 &  58855.56981879 & 236.1$\pm$4.5 & 24.94$\pm$2.89 & 5  & 12 \\
31 &  58855.57110841 & 239.3$\pm$2.2 & 14.01$\pm$1.49 & 7  & 14 \\
32 &  58855.57231400 & 243.0$\pm$2.5 & 13.70$\pm$1.56 & 6  & 12 \\
33 &  58855.57289030 & 241.2$\pm$2.6 & 14.30$\pm$1.73 & 6  & 12 \\
34 &  58855.57346368 & 241.7$\pm$6.3 & 23.11$\pm$4.65 & 3  & 8  \\
35 &  58855.57381872 & 241.9$\pm$1.9 & 20.17$\pm$1.25 & 10 & 23 \\
36 &  58855.58009888 & 239.4$\pm$2.5 & 11.75$\pm$1.58 & 6  & 10 \\
37 &  58855.58342643 & 239.5$\pm$2.5 & 12.78$\pm$1.72 & 6  & 11 \\
38 &  58855.58968159 & 239.1$\pm$4.7 & 24.03$\pm$3.20 & 4  & 11 \\
39 &  58855.59161939 & 240.1$\pm$2.0 & 11.78$\pm$1.29 & 7  & 13 \\
40 &  58855.59189638 & 240.5$\pm$3.1 & 18.71$\pm$1.98 & 6  & 13 \\
41 &  58855.59426110 & 240.5$\pm$1.0 & 20.60$\pm$1.19 & 20 & 46 \\
42 &  58855.59444177 & 241.1$\pm$0.9 & 11.56$\pm$0.56 & 16 & 28 \\

\hline
\specialrule{0.05em}{2pt}{2pt}
\end{tabular}

\setlength{\parskip}{-1.0ex}
\end{center}
\end{table}

\clearpage

\textbf{References}

\end{document}